\documentclass[sigconf,natbib=true]{acmart}

\usepackage{textcomp}
\usepackage[many]{tcolorbox} 
\usepackage{stfloats}
\usepackage{url}
\usepackage{verbatim}
\usepackage{amsthm}
\usepackage{subfigure}
\usepackage{makecell}
\usepackage{multirow}
\usepackage{enumitem}
\usepackage{subcaption}
\usepackage{algorithm}
\usepackage{algorithmic}

\usepackage{color, colortbl}

\definecolor{Gray}{rgb}{0.9, 0.9, 0.9}

\usepackage{float}

\usepackage{algorithm}
\usepackage{algorithmic}

\usepackage{color, colortbl}

\copyrightyear{2027}
\acmYear{2027}
\setcopyright{cc}
\setcctype{by}
\acmConference[WSDM '27]{Proceedings of the Twentieth ACM International Conference on Web Search and Data Mining}{February 15--19, 2027}{Hong Kong, Hong Kong}
\acmBooktitle{Proceedings of the Twentieth ACM International Conference on Web Search and Data Mining (WSDM '27), February 15--19, 2027, Hong Kong, Hong Kong}
\acmDOI{10.1145/3862322.3863720}
\acmISBN{979-8-4007-3117-4/2027/02}
\begin{document}

\title[DIGCN]{Take a Step Back: A Dual-Granularity Raw Feature Empowered Intermediate Fusion Graph Convolutional Network for Multimodal Recommendation}

\author{Jinfeng Xu}
\email{jinfeng.xu@ubc.ca}
\affiliation{%
  \institution{The University of British Columbia}
  \city{Vancouver}
  \country{Canada}}

\author{Zheyu Chen}
\email{zyueee.chen@connect.polyu.hk}
\affiliation{%
  \institution{The Hong Kong Polytechnic University}
  \city{Hong Kong}
  \country{Hong Kong}}

\author{Shuo Yang}
\email{shuoyang.ee@gmail.com}
\affiliation{%
  \institution{The University of Hong Kong}
  \city{Hong Kong}
  \country{Hong Kong}}

\author{Jinze Li}
\email{lijinze-hku@connect.hku.hk}
\affiliation{%
  \institution{The University of Hong Kong}
  \city{Hong Kong}
  \country{Hong Kong}}

\author{Wenhao Yuan}
\email{wenhao.yuan@connect.hku.hk}
\affiliation{%
  {\institution{The University of Hong Kong}}
  \city{Hong Kong}
  \country{Hong Kong}}

\author{Ziyue Peng}
\email{zpengbe@connect.ust.hk}
\affiliation{%
    \institution{The Hong Kong University of Science and Technology
}  
  \city{Hong Kong}
  \country{Hong Kong}}

\author{Puzhen Wu}
\email{puzhenwu8@connect.hku.hk}
\affiliation{%
  {\institution{The University of Hong Kong}}
  \city{Hong Kong}
  \country{Hong Kong}}

\author{Shujie Li}
\email{u3012850@connect.hku.hk}
\affiliation{%
  {\institution{The University of Hong Kong}}
  \city{Hong Kong}
  \country{Hong Kong}}

\author{Jian Chen}
\email{ccccccj03@connect.hku.hk}
\affiliation{%
  {\institution{The University of Hong Kong}}
  \city{Hong Kong}
  \country{Hong Kong}}

\author{Zheng Lin}
\email{zhenglin@ieee.org}
\affiliation{%
  \institution{University of Luxembourg}
  \city{Luxembourg}
  \country{Luxembourg}}

\author{Edith C. H. Ngai}
\authornote{Corresponding author.}
\email{chngai@hku.hk}
\affiliation{%
  {\institution{The University of Hong Kong}}
  \city{Hong Kong}
  \country{Hong Kong}}

\renewcommand{\shortauthors}{Jinfeng Xu et al.}




\renewcommand{\shortauthors}{Jinfeng Xu et al.}


\begin{abstract}
Although multimodal recommendations have shown the ability to learn user preferences and item properties by leveraging explicit multimodal information, they remain constrained by two critical limitations: low-quality representation learning and modality fusion. For the former, existing solutions primarily focus on multi-granularity representation learning; however, we argue that training such representations is difficult due to the inherent data sparsity problem in recommendation scenarios. For the latter, existing solutions often rely on contrastive learning to align representations of different modalities to facilitate modality fusion, yet alignment-only objectives may over-compress valuable modality-specific information. To this end, we suggest \underline{``Take a Step Back"} and propose a \textbf{D}ual-granularity raw feature empowered \textbf{I}ntermediate fusion \textbf{G}raph \textbf{C}onvolutional \textbf{N}etwork (\textbf{DIGCN}) for multimodal recommendation. DIGCN leverages the multimodal understanding capabilities of Large Language Models (LLMs) to construct dual-granularity raw features directly. Furthermore, DIGCN introduces an intermediate fusion graph convolutional network equipped with dynamic cross-modality weight correction, designed to capture cross-modality signals and achieve effective modality fusion. Comprehensive experiments on four real-world datasets validate that DIGCN consistently outperforms a variety of strong baselines. Additional analyses confirm its superiority under varying levels of data sparsity and cold-start scenarios. Notably, DIGCN demonstrates high extensibility, capable of further benefiting from additional contrastive learning-based alignment tasks.
\end{abstract}

\ccsdesc[500]{Information systems~Recommender systems;}



\keywords{Recommendation, Multimodal Learning}

\maketitle



\section{Introduction}
\label{sec:introduction}
The proliferation of online information has created an overload of content, making effective recommender systems critical to navigating this complexity \cite{xu2024aligngroup,chen2025squeeze}. Traditional recommender systems predominantly infer user preference through historical interaction data \cite{he2020lightgcn,xu2024fourierkan,xu2025nlgcl}. However, their performance is fundamentally limited by the data sparsity problem, as inferring comprehensive user and item profiles from sparse interactions remains a significant challenge. To mitigate this limitation, recent works have increasingly turned to multimodal recommender systems, leveraging rich auxiliary data from explicitly multimodal information to enrich representations and alleviate data sparsity problem \cite{xu2025vi,jiang2024diffmm,xu2025mdvt,xu2025enhancing}. 

Building upon multimodal data, existing multimodal recommender systems nevertheless face two persistent bottlenecks \cite{zhou2023comprehensive}: low-quality representation learning and modality fusion. \textbf{For representation learning}, current solutions to multi-granularity representation learning struggle to acquire sufficiently nuanced and discriminative features, primarily due to the intrinsic data sparsity in recommendation scenarios, which provides inadequate supervisory signal for training such complex representations \cite{chen2025hypercomplex,ren2025modality}. \textbf{For modality fusion}, current fusion strategies often rely on contrastive learning to align representations across modalities \cite{xu2025cohesion,wei2023multi,xu2025mentor}. While promoting consistency, alignment-only objectives may over-compress modality-specific information by minimizing inter-modal distance, thereby weakening unique, discriminative features that are crucial for recommendation. These two limitations, ineffective representation learning and information loss during fusion, work together to slow down further progress of multimodal recommendations. 
\begin{gather*}
    \textbf{Let us Take a Step Back!}
\end{gather*}
We therefore rethink these two limitations from a foundational perspective. First, if multi-granularity representation learning increases training difficulty, could we \textbf{take a step back} to construct multi-granularity raw features before training phase, thereby obtaining the desired representations directly? The powerful multimodal understanding capabilities of Large Language Models (LLMs) offer a feasible and potent means to realize this goal \cite{guo2025deepseek,yang2025qwen3}. Second, given that alignment-only multimodal objectives may weaken modality-specific information, could we \textbf{take a step back} to capture cross-modality signals before obtaining final representations via late fusion\footnote{According to a recent survey \cite{xu2025survey}, most advanced works adopt late fusion, integrating modalities after GCN.}? We use the message-passing stages in Graph Convolution Networks (GCNs) to realize this goal.

To this end, we propose a \textbf{D}ual-granularity raw feature empowered \textbf{I}ntermediate fusion \textbf{G}raph \textbf{C}onvolutional \textbf{N}etwork (\textbf{DIGCN}) for multimodal recommendation. Specifically, DIGCN leverages the powerful multimodal understanding capabilities of LLMs to construct dual-granularity raw features based on original multimodal information. Compared to multi-granularity representation learning, our dual-granularity raw feature component offers better interpretability while reducing the extra training difficulty. Moreover, DIGCN introduces an intermediate fusion graph convolutional network equipped with dynamic cross-modality weight correction, designed to capture cross-modality signals and achieve effective modality fusion. Compared to relying only on contrastive alignment, our intermediate fusion graph convolutional network preserves modality-specific information while fusing multiple modalities during propagation. Notably, DIGCN demonstrates high extensibility, capable of further benefiting from additional contrastive learning-based alignment tasks. Comprehensive experiments on four real-world datasets validate that DIGCN consistently outperforms a variety of strong baselines. Additional analyses confirm its superiority under varying levels of data sparsity and cold-start scenarios. In summary, our contributions include:
\begin{itemize}[leftmargin=*]
    \item We analyze the limitations of existing solutions for representation learning and modality fusion in multimodal recommendation.
    \item We propose DIGCN, a dual-granularity raw feature empowered intermediate fusion graph convolutional network for multimodal recommendation.
    \item We conduct comprehensive experiments on four real-world datasets, showing that DIGCN consistently outperforms a variety of strong baselines. Additional analyses confirm its superiority under varying levels of data sparsity and cold-start scenarios.
\end{itemize}

\section{Related Work}
With the growing diversity of information types on social media, multimodal information has been increasingly utilized to address the data sparsity problem in recommender systems \cite{zhou2023comprehensive,xu2025survey}. Early work, such as VBPR \cite{he2016vbpr}, incorporated visual content into matrix factorization \cite{rendle2012bpr} to alleviate sparsity issues. Subsequent works \cite{chen2019personalized,xu2025best,yu2023multi,chen2025don} extended this approach by integrating both visual and textual modalities, further enriching item representations and improving recommendation performance. Leveraging graph-based architectures, MMGCN \cite{wei2019mmgcn} pioneered the use of GCNs to extract modality-specific features from user-item interactions. Later models, such as DualGNN \cite{wang2021dualgnn} and LATTICE \cite{zhang2021mining}, introduced user-user and item-item graphs to better capture shared patterns in user preferences and item relationships. Building on these advancements, FREEDOM \cite{zhou2023tale} improved stability by freezing the item semantic graph and reducing noise in the user-item bipartite graph, while LOBSTER \cite{xu2025lobster} introduced global factors to capture cross-modal shared features.

Despite these advancements, multimodal recommender systems still face two major challenges: low-quality representation learning and modality fusion \cite{zhou2023comprehensive}. For representation learning, current approaches focus on multi-granularity representations. Specifically, MambaRec \cite{ren2025modality} constructs multi-scale representations to capture fine-grained semantic patterns, while HPMRec \cite{chen2025hypercomplex} employs hypercomplex embeddings with multi-components to enhance the diversity of multimodal features. For modality fusion, contrastive learning has become a common strategy for aligning multimodal representations. Specifically, MMSSL \cite{wei2023multi} and MICRO \cite{zhang2022latent} utilize contrastive learning to align multimodal signals with collaborative information, improving performance without extensive labeled data. BM3 \cite{zhou2023bootstrap} further explores inter-modal relationships to enhance recommendation accuracy, and MENTOR \cite{xu2025mentor} introduces a multi-level self-supervised task to align modalities without compromising user-item interactions. 

However, we argue that learning multi-granularity representations remains challenging due to inherent data sparsity, and alignment-only modality fusion may over-compress valuable modality-specific information. Therefore, we suggest to \textbf{take a step back} and propose DIGCN, which leverages large language models to construct dual-granularity raw features, avoiding extra training difficulty, and introduces an intermediate fusion GCN to achieve high-quality modality fusion.

\section{Methodology}

\begin{figure*}[h]
    \centering
    \includegraphics[width=1\linewidth]{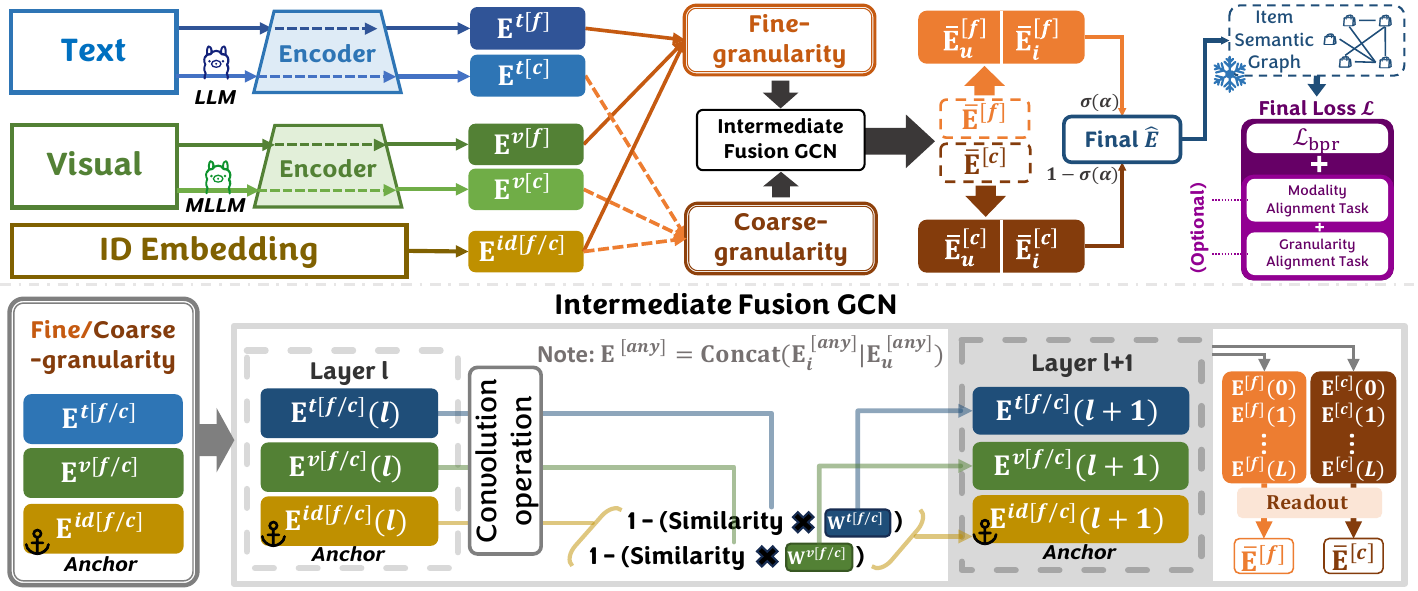}
    \vskip -0.1in
    \caption{Overall architecture of DIGCN, including dual-granularity raw feature construction, ID-anchored intermediate fusion, semantic item aggregation, and optional alignment tasks.}
    \label{fig:DIGCN}
\end{figure*}

Figure~\ref{fig:DIGCN} illustrates the overall structure of DIGCN\footnote{Code can be found in \href{https://github.com/Jinfeng-Xu/DIGCN}{https://github.com/Jinfeng-Xu/DIGCN}.}. DIGCN comprises three core components: (1) a dual-granularity raw feature construction module that leverages the multimodal understanding capabilities of MLLMs to generate coarse-grained raw features, which are combined with the original fine-grained raw features to form dual-granularity representations; (2) an intermediate fusion graph convolutional network equipped with dynamic cross-modality weight correction, which performs layer-wise modality fusion during the message-passing process to capture inter-modality features while preserving modality-specific features; and (3) an item-item semantic graph that enhances item representations through neighbor aggregation based on pre-computed feature similarities. We begin by introducing the preliminary notations and problem formulation in Section~\ref{sec:pre}, followed by the dual-granularity raw feature construction in Section~\ref{sec:dgrf}, the intermediate fusion GCN in Section~\ref{sec:ifgcn}, and the optimization objective in Section~\ref{sec:opti}.

\subsection{Preliminary}
\label{sec:pre}
Let $\mathcal{U}=\{u_1,...,u_{|\mathcal{U}|}\}$ denotes user set and $\mathcal{I}=\{i_1,...,i_{|\mathcal{I}|}\}$ denotes item set. We conceptualize the user-item graph $\mathcal{G} = (\mathcal{U}, \mathcal{I}, \mathcal{E})$, where $\mathcal{U}, \mathcal{I}$ serve as the graph vertices, and $\mathcal{E}$ denotes the edge set. In multimodal scenarios, each item contains multiple features. We introduce modality-specific item raw features $i^m$ for each item $i$ belonging to the set of modalities $\mathcal{M}$. For items, representations for each modality $m \in \mathcal{M}$ are extracted using a pre-trained extractor $\operatorname{Extractor}_m(\cdot)$, followed by a linear mapping $\phi(\cdot)$ to unify the dimensionality. Formally, this process is expressed as: $\mathbf{e}_{i}^{m} = \phi(\operatorname{Extractor}_m(i^m))$. For users, representations for each modality $m$ are aggregated from the item representations within the 1-hop training neighbor set, defined as: $\mathbf{e}_{u}^{m} = \sum_{i \in \mathcal{N}_{tr}(u)} \mathbf{e}_{i}^{m} / |\mathcal{N}_{tr}(u)|$, where $\mathcal{N}_{tr}(u)$ denotes the training neighbor set of user $u$. Additionally, the representations for the behavior modality $id$ of both users and items are randomly initialized. In summary, $\mathbf{E}_{i}^{m/id}$ and $\mathbf{E}_{u}^{m/id}$ represent the complete item and user representations for each modality $m/id$, respectively. The user-item interaction matrix is denoted as $\mathcal{R} \in \mathbb{R}^{|\mathcal{U}| \times |\mathcal{I}|}$. Specifically, each entry $\mathcal{R}_{u,i}$ indicates whether the user $u$ is connected to item $i$, with a value of $1$ representing a connection and $0$ otherwise.

\subsection{Dual-Granularity Raw Feature Construction}
\label{sec:dgrf}
Existing multi-granularity representation learning works \cite{chen2025hypercomplex,ren2025modality} in multimodal recommendation scenarios construct multiple representations based on $\mathbf{E}_{i}^{m/id}$ and $\mathbf{E}_{u}^{m/id}$, and use additional loss functions to provide self-supervised signals that distinguish different granularities. However, these designs significantly increase training difficulty. This issue becomes more challenging in recommendation scenarios, where users typically interact with only a limited number of items. Therefore, we suggest \textbf{take a step back} by constructing multi-granularity representations directly from raw features.

The rapid advancements in LLMs have enabled powerful multimodal understanding capabilities. We utilize LLMs to preprocess multimodal information and generate raw features at multiple granularities. However, constructing multiple raw features increases computational overhead, and the inherent randomness of LLMs may cause performance fluctuations. To ensure both efficiency and stability, we construct only one additional coarse-grained raw feature, which, combined with the original raw feature, forms dual-granularity raw features. As shown in Section~\ref{sec:extension}, while DIGCN benefits from incorporating additional granularities, this comes with reduced stability and increased computational overhead.

Specifically, for the textual modality, we utilize the prompt: \textit{``Please generate a coarse-grained version for this text {$i^t$}"}. For the visual modality, we use the prompt: \textit{``Please generate a coarse-grained version for this image {$i^v$} with this title {$i^t(\text{title})$}"}. In practice, the original product image is provided as the visual reference, and the item title is used as textual guidance to preserve the item identity while abstracting low-level visual details. Notably, following the MMRec setting, textual information is concatenated from the title, brand, categories, and description. Here, $i^t(\text{title})$ denotes the title information of item $i$. For clarity, we rename the original raw features $i^t$ and $i^v$ as $i^{t[f]}$ and $i^{v[f]}$, representing fine-grained raw features, and the generated raw features as $i^{t[c]}$ and $i^{v[c]}$, representing coarse-grained raw features. Accordingly, the relevant representations for modality $m$ are denoted as $\mathbf{e}_{u/i}^{m[f/c]} \in \mathbf{E}_{u/i}^{m[f/c]}$ for items and users. For the behavior modality $id$, we use one shared initial embedding and maintain branch-specific anchor states during fine- and coarse-grained propagation. Notably, in Section~\ref{sec:generation}, we further provide an analysis of replacing the original information with coarse-grained modality information constructed by LLMs for all baselines, to confirm that the advantages stem from dual granularity rather than the generated modality information.

\subsection{Intermediate Fusion GCN}
\label{sec:ifgcn}
Most recent multimodal recommendation studies \cite{zhang2021mining,zhou2023bootstrap,zhou2023comprehensive} adopt either early fusion, which integrates modalities before applying GCN, or late fusion, which combines modalities after GCN. However, early fusion may weaken intra-modality features during aggregation, while late fusion can miss cross-modality signals that emerge during graph propagation\footnote{For those unfamiliar with multimodal recommendation, see a recent survey for details on fusion strategies \cite{xu2025survey}.}. We argue that modality importance varies across different GCN layers. To address this limitation, we propose an intermediate fusion GCN equipped with dynamic cross-modality weight correction. Specifically, we treat the behavior modality $id$ as the anchor and fuse the representations of all other modalities during propagation. For compactness, let $g \in \{f,c\}$ denote the granularity branch, $x \in \mathcal{U}\cup\mathcal{I}$ denote either a user or an item, and $\mathcal{N}(\cdot)$ denote training neighbors in this subsection. At layer $l$, DIGCN first computes provisional anchor states and modality-specific propagated representations:
\begin{equation}
\begin{aligned}
    \tilde{\mathbf{e}}^{id[g]}_{x}(l)
    &=\sum_{\hat{x}\in\mathcal{N}(x)}
    \frac{\mathbf{e}^{id[g]}_{\hat{x}}(l-1)}
    {\sqrt{|\mathcal{N}(x)||\mathcal{N}(\hat{x})|}},\\[-0.5mm]
    \mathbf{e}_{x}^{m[g]}(l)
    &=\sum_{\hat{x}\in\mathcal{N}(x)}
    \frac{\mathbf{e}_{\hat{x}}^{m[g]}(l-1)}
    {\sqrt{|\mathcal{N}(x)||\mathcal{N}(\hat{x})|}} .
\end{aligned}
\end{equation}
When $x$ is an item, $\hat{x}$ indexes its neighboring users; when $x$ is a user, $\hat{x}$ indexes its neighboring items. The dynamic correction weight is then calculated from the provisional anchor and modality-specific representation:
\begin{equation}
 \mathbf{w}_{x}^{m[g]}(l)
 =1-\frac{(\mathbf{e}_{x}^{m[g]}(l))^{\top}\tilde{\mathbf{e}}^{id[g]}_{x}(l)}
 {\|\mathbf{e}_{x}^{m[g]}(l)\|\|\tilde{\mathbf{e}}^{id[g]}_{x}(l)\|}.
\end{equation}
Since the cosine similarity is bounded in $[-1,1]$, the correction weight is bounded in $[0,2]$; Eq.~(3) further normalizes all modality contributions through the denominator, preventing any single modality from dominating the fused anchor state.
Finally, DIGCN fuses the modality-specific representations into the anchor state:
\begin{equation}
    \mathbf{e}^{id[g]}_{x}(l)
    =\frac{\tilde{\mathbf{e}}^{id[g]}_{x}(l)+\sum_{m\in\mathcal{M}}\mathbf{w}_x^{m[g]}(l)\mathbf{e}^{m[g]}_{x}(l)}
    {1+\sum_{m\in\mathcal{M}}\mathbf{w}_x^{m[g]}(l)}.
\end{equation}

Notably, compared to other multimodal recommendation works, our intermediate fusion GCN does not introduce significant computational overhead, as it essentially performs graph message-passing operation for each modality while fusing modalities during the process. In practice, the pre-constructed dual-granularity representations and intermediate fusion enable DIGCN to achieve competitive efficiency. We provide empirical evidence in Section~\ref{sec:efficiency}. We redefine $\mathbf{e}_{u/i}^{[f/c]}(l) = \mathbf{e}_{u/i}^{id[f/c]}(l)$ to better describe the representation aggregated from all modalities at the $l$-th layer with $id$ as the anchor. The final representations for all granularities are obtained via readout function $\operatorname{Readout}(\cdot)$:
\begin{equation}
    \mathbf{\bar{e}}_{u/i}^{[f/c]} = \operatorname{Readout}(\mathbf{e}_{u/i}^{[f/c]}(1),...,\mathbf{e}_{u/i}^{[f/c]}(L)),
\end{equation}
where the $\operatorname{Readout}(\cdot)$ can be any differentiable function, we adopt the mean function in practice. Then, we aggregate the representations of all granularities using a learnable weight:
\begin{equation}
    \mathbf{\bar{e}}_{u/i} = \sigma(\alpha) \mathbf{\bar{e}}_{u/i}^{[f]} + (1 - \sigma(\alpha))\mathbf{\bar{e}}_{u/i}^{[c]},
\end{equation}
where $\sigma(\cdot)$ denotes the Sigmoid function, and $\alpha$ is a learnable weight initialized to $0$, ensuring that $\sigma(\alpha)$ starts at $0.5$. Most advanced multimodal recommendation works \cite{xu2025survey,zhou2023comprehensive} adopt additional homogeneous graphs to enhance representations. In contrast, our DIGCN utilizes only a single item-item graph to enhance representations while minimizing computational overhead. To further reduce computational costs, we pre-construct the item-item graph based on the similarity of item raw features, and this graph remains frozen during training. The pairwise similarity based on raw features between all items is calculated as follows:
\begin{equation}
    \mathbf{S}_{p,q}=\sum_{m \in \mathcal{M}}\frac{(\mathbf{i}_{p}^m)^{\top} \mathbf{i}_{q}^m}{\|\mathbf{i}_{p}^m\|\|\mathbf{i}_{q}^m\|},
\end{equation}
we retain only the top-$k$ neighbors with the highest similarity scores to aggregate the most relevant neighbors:
\begin{equation}
\hat{\mathbf{S}}_{p,q} = \begin{cases}1 & \text { if } \mathbf{S}_{p,q} \in \text { top-} k(\mathbf{S}_{p,*}) \\ 0 & \text { otherwise }\end{cases}.
\end{equation}

To mitigate the issues of gradient explosion or vanishing, we normalize the similarity adjacency matrices as follows: $\mathbf{\tilde{S}}=\mathcal{D}^{-\frac{1}{2}} \hat{\mathbf{S}}\mathcal{D}^{-\frac{1}{2}}$, where $\mathcal{D}$ is the diagonal degree matrix of $\hat{\mathbf{S}}$. Let $\bar{\mathbf{E}}_{\mathcal{I}} \in \mathbb{R}^{|\mathcal{I}|\times d}$ denote the row-stacked item representation matrix after granularity aggregation. Then, we aggregate one-hop neighbors to enhance item representations:
\begin{equation}
    \bar{\mathbf{E}}_{\mathcal{I}} \leftarrow \bar{\mathbf{E}}_{\mathcal{I}} + \mathbf{\tilde{S}}\bar{\mathbf{E}}_{\mathcal{I}}.
\end{equation}

To preserve the personalization of each item's representation, we add the enhanced representation back to the original representation. This strategy ensures that each item's unique properties are maintained while benefiting from the aggregation of its neighbors' representations. Notably, unlike most works constructing $|\mathcal{M}|$ item-item graphs and aggregating multi-hop neighbors, DIGCN leverages fine-grained representations and high-quality modality fusion, requiring only a single item-item graph and one-hop aggregation, thus reducing computational overhead.

\subsection{Optimization}
\label{sec:opti}
We adopt the Bayesian Personalized Ranking (BPR) \cite{rendle2012bpr} loss as the optimization function. BPR aims to maximize the preference divergence between positive and negative items within each user-item triplet $(u, p, n) \in \mathcal{D}$, where $\mathcal{D}$ signifies the collection of training data, the term positive item $p$ pertains to an item with which the user $u$ has interacted, whereas the negative item $n$ is selected randomly from the pool of items without interaction with user $u$.
\begin{equation}
\mathcal{L}_{bpr}=\sum_{(u, p, n) \in \mathcal{D}} - \ln \sigma\left(\hat{y}_{u,p}-\hat{y}_{u,n}\right)+\lambda\left\|\mathbf{\Theta}\right\|^2,
\end{equation}
where $\lambda$ controls the L2 regularization, $\sigma(\cdot)$ is the Sigmoid function, and $\mathbf{\Theta}$ denotes model parameters. $\hat{y}_{u,p}$ and $\hat{y}_{u,n}$ are calculated by $(\mathbf{\bar{e}}_{u})^{\top}\mathbf{\bar{e}}_{p}$ and $(\mathbf{\bar{e}}_{u})^{\top} \mathbf{\bar{e}}_{n}$, respectively. Notably, DIGCN demonstrates high extensibility and can further benefit from additional contrastive alignment tasks, as empirically analyzed in Section~\ref{sec:compatibility}.

\begin{table}[!t]
    \centering
    \caption{Statistics of the experimental datasets.}
    \small
    \label{tab:dataset_statistics}
    \begin{tabular}{ccccc}
    \toprule
         Dataset&  \# Users&  \# Items&  \# Interaction& Sparsity\\
         \midrule
         Baby &  19445 &  7050 &  160792 & 99.88\%\\
         Sports &  35598 &  18357 &  296337 & 99.95\%\\
         Pet &  19856 &  8510 &  157836 & 99.91\%\\
         Office &  4905 &  2420 &  53258 & 99.55\%\\
         \bottomrule
    \end{tabular}
\end{table}

\section{Experiment}
\label{sec:experiment}

\subsection{Experiment Setup}
\subsubsection{Datasets}
To evaluate the effectiveness of our DIGCN, we conduct extensive experiments on four widely used Amazon review datasets~\cite{mcauley2015image}: (a) Baby, (b) Sports and Outdoors (Sports for short), (c) Pet Supplies (Pet for short), and (d) Office Products (Office for short). These datasets span diverse product domains with varying scales of users, items, and interactions, providing a comprehensive testbed for evaluating multimodal recommendation systems. Each dataset contains rich multimodal information, including product images and textual descriptions (comprising title, brand, categories, and description fields), which serve as the visual and textual modalities in our framework. Following most previous works~\cite{zhou2023bootstrap,wang2021dualgnn}, we adopt the 5-core setting to filter out users and items with fewer than five interactions, ensuring that each user and item has a minimum level of interaction history for meaningful evaluation. For each dataset, we randomly split the historical interactions into training, validation, and testing sets using an 8:1:1 ratio. The key statistics of these datasets after preprocessing are delineated in Table~\ref{tab:dataset_statistics}. 

\begin{table*}[!t]
    \centering
    \caption{Overall performance over five random seeds. $^*$ denotes statistical significance with p-value $<$ $0.05$.}
    \label{tab:result}
    \resizebox{\linewidth}{!}{
    \begin{tabular}{c|cccccccccccccccc}
    \toprule
    \multicolumn{1}{c}{Datasets}&  \multicolumn{4}{c}{Baby}&  \multicolumn{4}{c}{Sports}&  \multicolumn{4}{c}{Pet}& \multicolumn{4}{c}{Office}\\\midrule 
    \multicolumn{1}{c}{Metrics}& R@10& R@20& N@10& N@20& R@10& R@20& N@10& N@20& R@10& R@20& N@10& N@20& R@10& R@20& N@10& N@20 \\\midrule
    VBPR & 0.0423 &0.0663& 0.0223 &0.0284& 0.0558 &0.0856& 0.0307 &0.0384 & 0.0928 &0.1278& 0.0469 &0.0598& 0.0692 &0.1084& 0.0422 &0.0531\\
    MMGCN & 0.0378 &0.0615& 0.0200 &0.0261& 0.0370 &0.0605& 0.0193 &0.0254 & 0.0679 &0.1002& 0.0362 &0.0440& 0.0558 & 0.0926& 0.0312 &0.0413\\
    DualGNN & 0.0448 &0.0716& 0.0240 &0.0309& 0.0568 &0.0859& 0.0310 &0.0385 & 0.0965 &0.1347& 0.0491 &0.0620& 0.0887 &0.1350& 0.0505 &0.0631\\
    LATTICE & 0.0547 &0.0850& 0.0292 &0.0370& 0.0620 &0.0953& 0.0335 &0.0421 & 0.1029 &0.1449& 0.0532 &0.0675& 0.0969 &0.1421& \underline{0.0562} &\underline{0.0686}\\
    FREEDOM & 0.0627& 0.0992& 0.0330& 0.0424& 0.0717& \underline{0.1089}& 0.0385 & \underline{0.0481} & \underline{0.1135} & \underline{0.1643}& \underline{0.0587} & \underline{0.0755}& 0.0974& 0.1445& 0.0549&0.0669\\
    SLMRec & 0.0529 &0.0775& 0.0290 &0.0353& 0.0663 &0.0990& 0.0365 &0.0450 & 0.1043 &0.1476& 0.0547 &0.0692& 0.0790 &0.1252& 0.0475 &0.0599\\
    BM3 & 0.0564 &0.0883& 0.0301 &0.0383& 0.0656 &0.0980& 0.0355 &0.0438 & 0.0995 &0.1384& 0.0494 &0.0625& 0.0715 &0.1155& 0.0415 &0.0533\\
    MMSSL & 0.0613 &0.0971& 0.0326 &0.0420& 0.0693 &0.1013& 0.0369 &0.0474 & 0.1069 &0.1502& 0.0555 &0.0706& 0.0794 &0.1273& 0.0481 &0.0610\\
    LLMRec & 0.0621 & 0.0983 & 0.0324 & 0.0422 & 0.0682 & 0.1000 & 0.0363 & 0.0459 & 0.1082 & 0.1531 & 0.0561 & 0.0713 & 0.0809 & 0.1299 & 0.0492 & 0.0621 \\
    LGMRec & 0.0639& 0.0989& 0.0337& 0.0430& 0.0719& 0.1068& \underline{0.0387}& 0.0477& 0.1057& 0.1545 & 0.0584& 0.0737& 0.0959& 0.1402& 0.0514& 0.0663\\
    DiffMM & 0.0623& 0.0975& 0.0328& 0.0411& 0.0671& 0.1017& 0.0377& 0.0458& 0.1055& 0.1494& 0.0551& 0.0700& 0.0733& 0.1183& 0.0439& 0.0560\\
    LOBSTER & 0.0563 & 0.0870& 0.0296& 0.0375& 0.0594& 0.0913& 0.0318& 0.0400& 0.0991& 0.1401& 0.0490& 0.0629& 0.0719& 0.1149& 0.0419& 0.0530\\
    BeFA & 0.0555& 0.0884& 0.0299& 0.0383& 0.0649& 0.0985& 0.0346& 0.0432& 0.0998& 0.1399& 0.0501& 0.0631& 0.0732& 0.1171& 0.0431& 0.0549\\ 
    MambaRec& \underline{0.0650}& 0.1013& \underline{0.0353}& \underline{0.0444}& \underline{0.0723}& 0.1084& 0.0385& 0.0480& 0.1107& 0.1589& 0.0580& 0.0748& \underline{0.0989}& \underline{0.1449}& 0.0559& 0.0683\\
    DHMRec& 0.0646& \underline{0.1016}& 0.0343& 0.0436& 0.0721& 0.1082& 0.0382& 0.0478& 0.1115& 0.1603& 0.0583& 0.0752& 0.0982& 0.1441& 0.0555& 0.0680\\\midrule
    DIGCN & \textbf{0.0686$^*$}&\textbf{0.1052$^*$}& \textbf{0.0363$^*$} &\textbf{0.0459$^*$}& \textbf{0.0765$^*$} &\textbf{0.1140$^*$}& \textbf{0.0411$^*$} &\textbf{0.0512$^*$}& \textbf{0.1214$^*$}&\textbf{0.1729$^*$}& \textbf{0.0634$^*$} &\textbf{0.0799$^*$}& \textbf{0.1040$^*$}&\textbf{0.1543$^*$}& \textbf{0.0602$^*$}&\textbf{0.0740$^*$} \\ \bottomrule
    \end{tabular}
    }
\end{table*}

\subsubsection{Evaluation Metric}
For a fair and comprehensive comparison, we adopt two widely-used ranking-based evaluation metrics: Recall and Normalized Discounted Cumulative Gain (NDCG). We report the average metrics computed over all users in the test set under four settings: Recall@10 (R@10), Recall@20 (R@20), NDCG@10 (N@10), and NDCG@20 (N@20). 

\subsubsection{Baseline}
To verify the effectiveness of DIGCN, we compare it against the following multimodal recommendation models, including \textbf{VBPR} \cite{he2016vbpr}, \textbf{MMGCN} \cite{wei2019mmgcn}, \textbf{DualGNN} \cite{wang2021dualgnn}, \textbf{LATTICE} \cite{zhang2021mining}, \textbf{FREEDOM} \cite{zhou2023tale}, \textbf{SLMRec} \cite{tao2022self}, \textbf{BM3} \cite{zhou2023bootstrap}, \textbf{MMSSL} \cite{wei2023multi}, \textbf{LLMRec} \cite{wei2024llmrec}, \textbf{LGMRec} \cite{guo2024lgmrec}, \textbf{DiffMM} \cite{jiang2024diffmm}, \textbf{LOBSTER} \cite{xu2025lobster}, \textbf{BeFA} \cite{fan2025befa}, \textbf{MambaRec} \cite{ren2025modality}, and \textbf{DHMRec} \cite{zhan2026dhmrec}.

\subsubsection{Implementation Detail}
We implement our proposed DIGCN and all the baselines using MMRec \cite{zhou2023mmrecsm}. Specifically, we chose Xavier initialization \cite{glorot2010understanding} of dimension $64$ for all initial random representation. Besides, we choose Adam optimizer \cite{kingma2014adam} for model training. Notably, we perform a complete grid search for all baselines to find the optimal setting based on their published papers. As for hyper-parameter settings on DIGCN, we perform a grid search on the layer number $L$ of Intermediate Fusion GCN in $\{1, 2, 3, 4\}$ and $k$ of top-$k$ in the item-item graph is searched from $\{10, 20, 30, 40\}$. We fix the learning rate with $10^{-4}$. To avoid the over-fitting problem, we set $20$ as the early stopping epoch number. Following previous work \cite{zhou2023mmrecsm}, we utilize Recall@20 on the validation dataset as a metric to update the best record. For dual-granularity raw feature construction, we use GPT-4o to generate coarse-grained textual descriptions and DALL-E-3 to generate coarse-grained visual references. Additionally, we provide relevant experiments with different LLMs in Section~\ref{sec:llms}.

\subsection{Overall Performance}
\label{sec:overall}
We evaluate DIGCN on four widely-used datasets against all baseline methods. Table~\ref{tab:result} leads to the following observations:
\begin{itemize}[leftmargin=*]
    \item DIGCN achieves statistically significant gains over all baselines on all datasets and metrics. Compared with the strongest baseline MambaRec, its relative Recall@20 improvements are 3.54\%, 5.17\%, 8.81\%, and 6.49\% on Baby, Sports, Pet, and Office, respectively, supporting the usefulness of dual-granularity raw features and ID-anchored intermediate fusion.
    \item Contrastive-learning baselines (e.g., MMSSL, BM3, and MambaRec) remain competitive, but DIGCN performs better without requiring an alignment loss. This suggests that intermediate fusion provides a complementary integration mechanism; Section~\ref{sec:compatibility} further studies optional alignment tasks. The larger gains on Pet and Office may be related to their more verbose product descriptions, where coarse-grained abstraction can better preserve core item semantics.
\end{itemize}

\subsection{Ablation Study}
\label{sec:ablation}
To validate the contribution of each component, we compare DIGCN with four variants: w/o Dual-granularity removes coarse-grained raw features, w/o Intermediate fusion GCN replaces intermediate fusion with late fusion, w/o Item-item graph removes semantic neighbor aggregation, and Uniform Weight keeps the fusion architecture but uses fixed modality weights.

\begin{figure}[!t]
    \centering
    \includegraphics[width=1\linewidth]{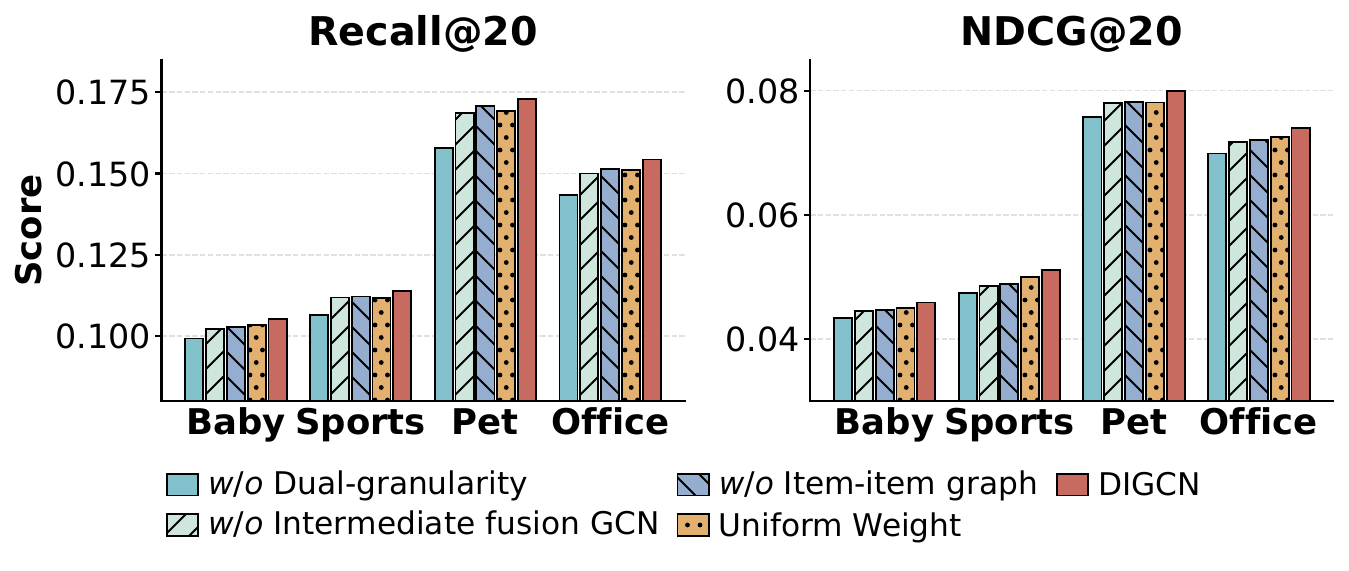}
    \caption{Ablation study across datasets.}
    \label{fig:ablation}
\end{figure}

As shown in Figure~\ref{fig:ablation}, DIGCN consistently performs best, while each removal causes a clear drop. The dual-granularity component is especially helpful on Pet and Office, where product descriptions are more verbose; the intermediate fusion GCN confirms the value of exchanging cross-modality signals during propagation; Uniform Weight shows that dynamic correction contributes beyond the fusion structure itself; and the item-item graph provides useful supplementary semantic neighbors.

\subsection{Compatibility Analysis}
\label{sec:compatibility}
To test whether DIGCN can benefit from additional learning objectives, we add three InfoNCE-based alignment variants following previous works~\cite{zhou2023bootstrap,ren2025modality}: +M-Align aligns textual and visual representations, +G-Align aligns fine- and coarse-grained representations, and +MG-Align combines both. The optimal temperature $\tau$ is consistently $0.2$ across datasets. Table~\ref{tab:compatibility} shows that all alignment variants improve the base DIGCN. +G-Align is usually slightly stronger than +M-Align, and +MG-Align performs best, indicating that modality and granularity alignment capture complementary coherence signals. Since the base DIGCN already outperforms contrastive baselines, these tasks are best viewed as optional enhancements rather than required components.

\begin{table}[!t]
    \centering
    \caption{Compatibility analysis of alignment tasks.}
    \label{tab:compatibility}
    \resizebox{\linewidth}{!}{
    \begin{tabular}{c|cccccccc}
    \toprule
    \multicolumn{1}{c}{Datasets}&  \multicolumn{2}{c}{Baby}&  \multicolumn{2}{c}{Sports}&  \multicolumn{2}{c}{Pet}& \multicolumn{2}{c}{Office}\\\midrule 
    \multicolumn{1}{c}{Metrics}& R@20& N@20& R@20& N@20& R@20& N@20& R@20& N@20 \\\midrule
    DIGCN & 0.1052& 0.0459& 0.1140& 0.0512& 0.1729& 0.0799& 0.1543& 0.0740 \\ 
    +M-Align & 0.1058& 0.0462& 0.1148& 0.0516& 0.1735& 0.0802& 0.1551& 0.0745 \\
    +G-Align & 0.1059& 0.0464& 0.1153& 0.0519& 0.1740& 0.0807& 0.1560& 0.0748 \\
    +MG-Align & 0.1063& 0.0466& 0.1158& 0.0522& 0.1750& 0.0812& 0.1569& 0.0754 \\\bottomrule
    \end{tabular}
    }
\end{table}

\begin{figure*}[t]
    \centering
    \includegraphics[width=1\linewidth]{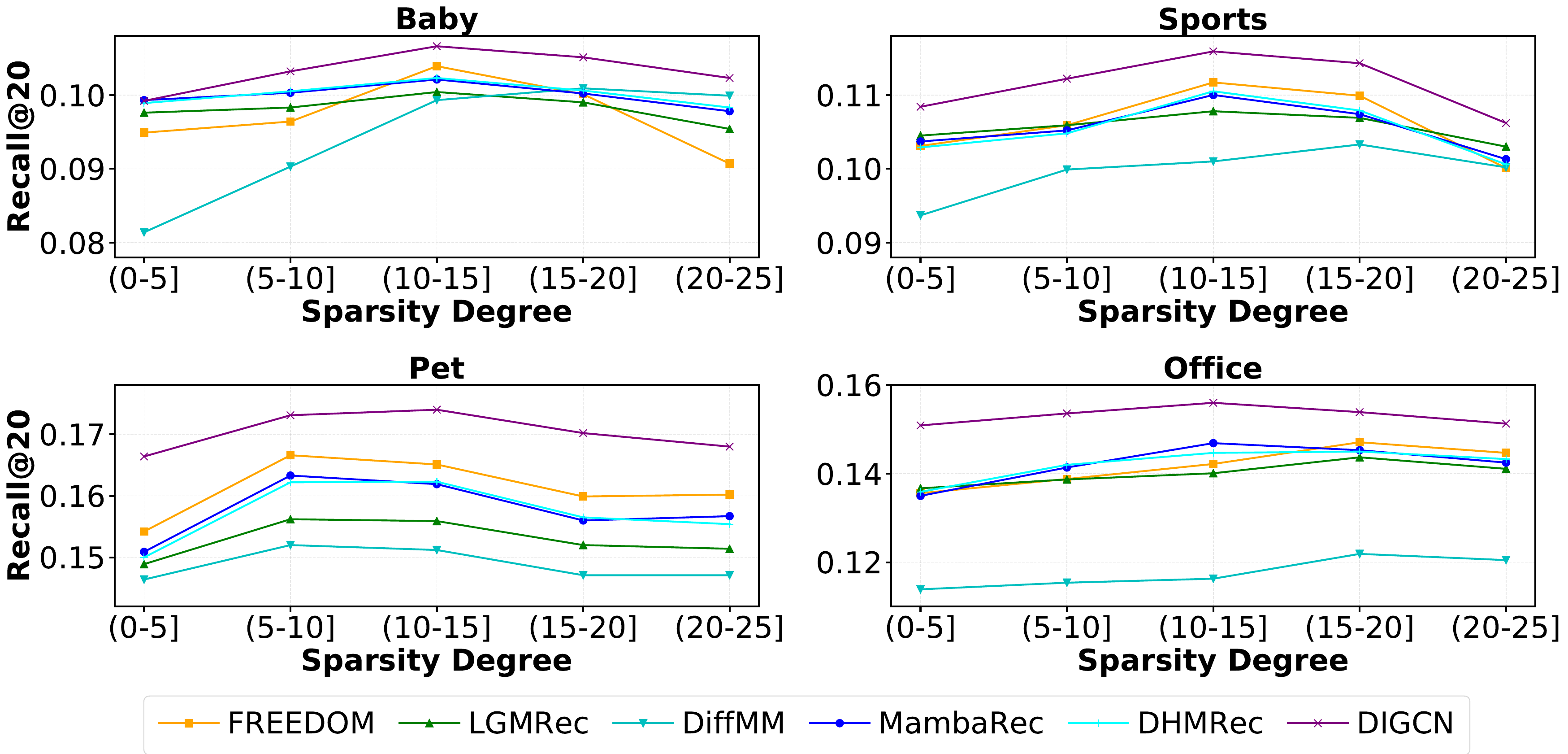}
    \caption{Sparsity analysis across datasets.}
    \label{fig:sparsity}
\end{figure*}

\subsection{Sparsity Analysis}
\label{sec:sparsity}
To evaluate robustness under data sparsity, we group users by the number of training interactions: $(0-5]$, $(5-10]$, $(10-15]$, $(15-20]$, and $(20-25]$. We compare DIGCN with competitive baselines, including FREEDOM, LGMRec, DiffMM, MambaRec, and DHMRec. As shown in Figure~\ref{fig:sparsity}, DIGCN consistently performs best across sparsity groups, with larger gains in the sparsest groups $(0-5]$ and $(5-10]$. This supports our motivation that pre-constructed dual-granularity features can provide useful item semantics when collaborative signals are limited, while intermediate fusion helps exploit the scarce interactions more effectively.

\begin{table}[!t]
    \centering
    \caption{Item cold-start performance.}
    \label{tab:cold-start}
    \resizebox{\linewidth}{!}{
    \begin{tabular}{l|cccccccc}
    \toprule
          \multicolumn{1}{c}{Datasets}& \multicolumn{2}{c}{Baby}&  \multicolumn{2}{c}{Sports}& \multicolumn{2}{c}{Pet}& \multicolumn{2}{c}{Office}\\\midrule
          \multicolumn{1}{c}{Metrics}& R@20& N@20& R@20& N@20& R@20& N@20& R@20& N@20 \\\midrule
          VBPR& 0.0192& 0.0107& 0.0234& 0.0129& 0.0350& 0.0172& 0.0320& 0.0156\\
          MMGCN& 0.0186& 0.0098& 0.0178& 0.0094& 0.0298& 0.0148& 0.0270& 0.0134\\
          DualGNN& 0.0200& 0.0110& 0.0242& 0.0132& 0.0374& 0.0183& 0.0371& 0.0178\\
          LATTICE& 0.0256& 0.0138& 0.0340& 0.0189& 0.0410& 0.0208& 0.0471& 0.0249\\
          FREEDOM& 0.0588& 0.0257& 0.0640& 0.0289& 0.0513& 0.0264& 0.0481& 0.0250\\
          SLMRec& 0.0259& 0.0140& 0.0354& 0.0194& 0.0419& 0.0215& 0.0365& 0.0174\\
          BM3& 0.0262& 0.0133& 0.0294& 0.0180& 0.0401& 0.0203& 0.0349& 0.0166\\
          MMSSL& 0.0351& 0.0192& 0.0370& 0.0203& 0.0441& 0.0226& 0.0369& 0.0174\\
          LLMRec& 0.0605& 0.0255& 0.0600& 0.0261& 0.0459& 0.0235& 0.0374& 0.0178\\
          LGMRec& 0.0592& 0.0261& 0.0629& 0.0281& 0.0461& 0.0237& 0.0460& 0.0241\\
          DiffMM& 0.0552& 0.0238& 0.0589& 0.0254& 0.0452& 0.0230& 0.0412& 0.0195\\
          LOBSTER& 0.0280& 0.0162& 0.0320& 0.0190& 0.0408& 0.0209& 0.0409& 0.0191\\
          BeFA& 0.0262& 0.0149& 0.0303& 0.0182& 0.0399& 0.0202& 0.0419& 0.0189\\
          MambaRec& 0.0611& 0.0252& 0.0633& 0.0285& 0.0497& 0.0255& 0.0486& 0.0255\\
          DHMRec& 0.0594& 0.0241& 0.0611& 0.0274& 0.0488& 0.0251& 0.0482& 0.0253\\
          DIGCN& \textbf{0.0633}& \textbf{0.0265}& \textbf{0.0669}& \textbf{0.0303}& \textbf{0.0552}& \textbf{0.0290}& \textbf{0.0544}& \textbf{0.0291}\\
    \bottomrule
    \end{tabular}
    }
\end{table}

\subsection{Cold-start Analysis}
\label{sec:cold-start}
We report item cold-start results following the widely used setting in \cite{zhang2022latent}: 20\% of items are randomly selected, all their interactions are removed from training, and the held-out items are split equally into validation and testing sets.

As shown in Table~\ref{tab:cold-start}, DIGCN consistently outperforms all baselines, achieving Recall@20 of $0.0633$, $0.0669$, $0.0552$, and $0.0544$ on Baby, Sports, Pet, and Office, respectively. The gain is consistent with the design of DIGCN: dual-granularity raw features provide content semantics for unseen items, and intermediate fusion builds more complete item profiles from visual and textual information.

\begin{figure}[t]
    \centering
    \includegraphics[width=1\linewidth]{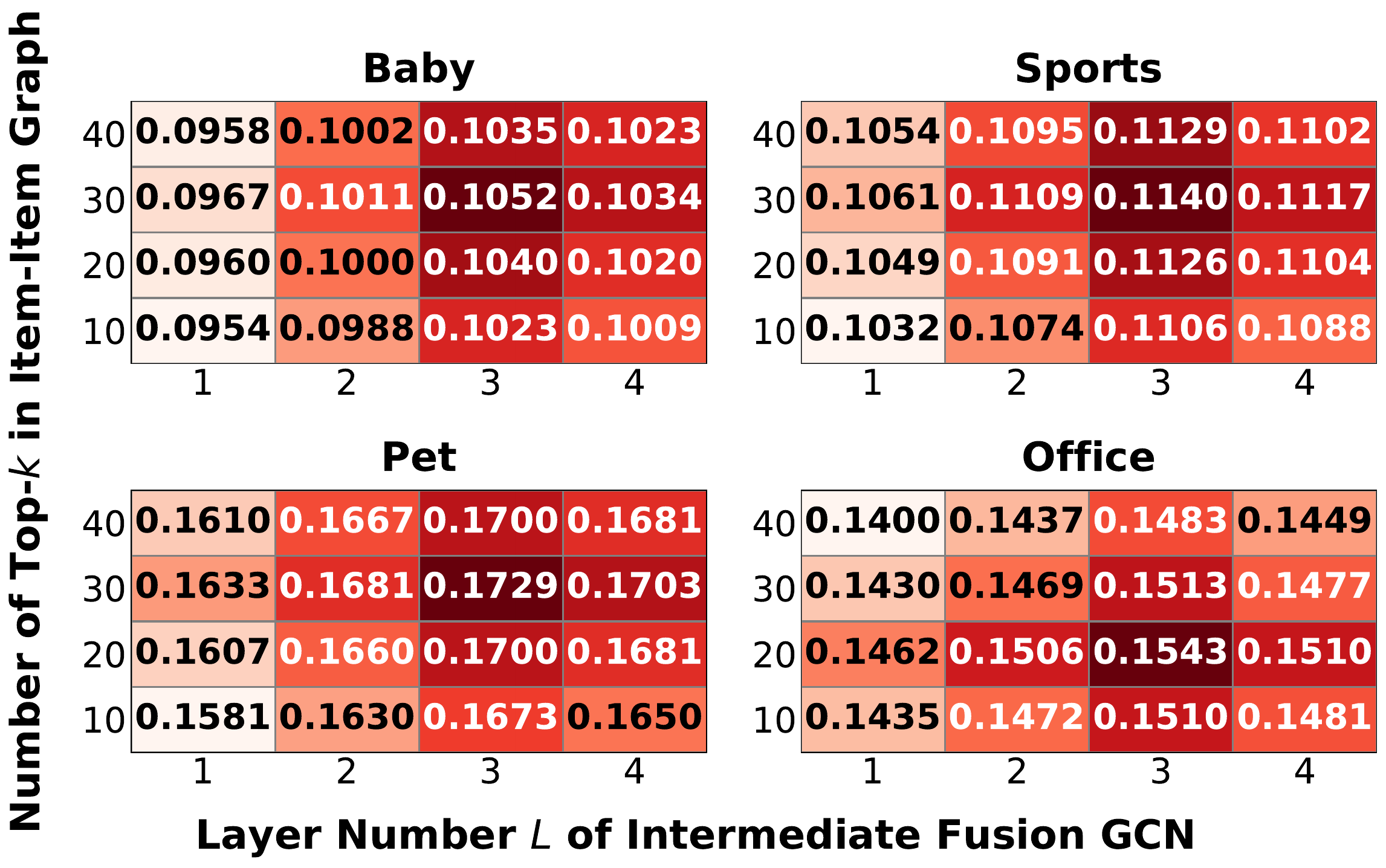}
    \caption{Hyperparameter sensitivity of DIGCN.}
    \label{fig:hyper-parameter}
\end{figure}

\subsection{Hyper-parameter Analysis}
\label{sec:hyper-parameter}
We study the layer number $L$ of the intermediate fusion GCN and the top-$k$ neighbors in the item-item graph. As shown in Figure~\ref{fig:hyper-parameter}, $L=3$ performs best across datasets; shallower settings may underuse message passing, while $L=4$ can introduce over-smoothing. The best $k$ is $30$ on Baby, Sports, and Pet, and $20$ on Office, whose item set is smaller. Overall, DIGCN is stable within the tested ranges and does not require extensive dataset-specific tuning.

\begin{table}[!t]
    \centering
    \caption{LLM sensitivity on Recall@20. Std. is computed over five prompt rewrites.}
    \label{tab:LLM}
    \resizebox{\linewidth}{!}{
    \begin{tabular}{c|cccc}
    \toprule
        \multicolumn{1}{c}{Datasets}&  \multicolumn{2}{c}{Baby}&  \multicolumn{2}{c}{Sports}\\\midrule
        Metrics& Mean& Std.& Mean& Std.\\ \midrule
        \multicolumn{5}{c}{Close Source}\\\midrule
        GPT-4o + DALL-E-3& 0.1052& 0.0009& 0.1140& 0.0007\\
        Qwen3-Max + DALL-E-3& 0.1059& 0.0012& 0.1153& 0.0004\\
        Gemini-2.5-Pro + DALL-E-3& 0.1055& 0.0006& 0.1149& 0.0009\\
        GPT-4o + FLUX1.1& 0.1054& 0.0010& 0.1143& 0.0013\\
        Qwen3-Max + FLUX1.1& 0.1058& 0.0009& 0.1155& 0.0008\\
        Gemini-2.5-Pro + FLUX1.1& 0.1054& 0.0008& 0.1150& 0.0004\\
        GPT-4o + MidJourney v7& 0.1059& 0.0010& 0.1151& 0.0008\\
        Qwen3-Max + MidJourney v7& 0.1061& 0.0011& 0.1158& 0.0013\\
        Gemini-2.5-Pro + MidJourney v7& 0.1060& 0.0009& 0.1160& 0.0009\\\midrule
        \multicolumn{5}{c}{Open Source}\\\midrule
        DeepSeek-R1 + Diffusion 3.5& 0.1056& 0.0008& 0.1150& 0.0010\\
    \bottomrule
    \end{tabular}
    }
\end{table}

\subsection{Analysis for Different LLMs}
\label{sec:llms}
We examine whether DIGCN depends on specific LLM choices. For coarse-grained text generation, we compare GPT-4o, Qwen3-Max, Gemini-2.5-Pro, and DeepSeek-R1; for coarse-grained visual generation, we compare DALL-E-3, FLUX1.1, MidJourney v7, and Stable Diffusion 3.5, forming ten text-image combinations. We also rewrite prompts five times with unchanged semantics and report the mean and standard deviation. Table~\ref{tab:LLM} shows that stronger LLM combinations bring small additional gains, while the standard deviations remain low. The open-source combination is also competitive, suggesting that DIGCN is not tightly coupled to a specific commercial generator or prompt wording.

\begin{table}[!t]
    \centering
    \caption{Efficiency comparison on Baby and Sports. Time is s/epoch, memory is GB, and R@20 is Recall@20.}
    \label{tab:efficiency}
    \resizebox{\linewidth}{!}{
    \begin{tabular}{c|cccccc}
    \toprule
        \multicolumn{1}{c}{Datasets}&  \multicolumn{3}{c}{Baby}&  \multicolumn{3}{c}{Sports}\\\midrule
        Metrics& Time& Memory& R@20& Time& Memory &R@20\\ \midrule
        LATTICE& 3.20& 4.53& 0.0850& 11.07& 19.93& 0.0953\\ 
        FREEDOM& 2.57& 2.13& 0.0992& 5.65& 3.34& 0.1089\\ 
        MMSSL& 6.31& 3.77& 0.0971& 14.67& 5.34& 0.1013\\ 
        LLMRec& 7.58& 4.04& 0.0983& 14.92& 5.19& 0.1000\\ 
        LGMRec& 4.19& 2.41& 0.0989& 8.39& 3.67& 0.1068\\ 
        DiffMM& 9.45& 4.23& 0.0975& 18.61& 5.99& 0.1017\\ 
        LOBSTER& 3.13& 2.81& 0.0870& 7.22& 4.10& 0.0913\\ 
        BeFA& 5.30& 3.89& 0.0884& 10.05& 5.21& 0.0985\\ 
        MambaRec& 9.01& 4.88& 0.1013& 15.39& 6.54& 0.1084\\
        DHMRec& 4.45& 4.12& 0.1016& 9.91& 5.87& 0.1082\\
        \midrule
        DIGCN& 2.99& 2.40& 0.1052& 6.91& 3.85& 0.1140\\ 
    \bottomrule
    \end{tabular}
    }
\end{table}

\subsection{Efficiency Analysis}
\label{sec:efficiency}

Table~\ref{tab:efficiency} compares training time and GPU memory on Baby and Sports. DIGCN achieves the best recommendation performance with competitive resource usage because LLM generation is an offline preprocessing step, the semantic graph is unified rather than modality-specific, and item-item aggregation is one-hop. Once coarse-grained features are generated, they can be reused across training runs; the LLM cost is therefore incurred only once, with $|\mathcal{I}|$ calls per modality.

\begin{table}[!t]
    \centering
    \small
    \caption{Multi-granularity extension on Recall@20. Std. is computed over five generations.}
    \label{tab:extension}
    \begin{tabular}{c|cccc}
    \toprule
        \multicolumn{1}{c}{Datasets}&  \multicolumn{2}{c}{Baby}&  \multicolumn{2}{c}{Sports}\\\midrule
        Metrics& Mean& Std.& Mean& Std.\\ \midrule
        Original& 0.1052& 0.0009& 0.1140& 0.0007\\
        3 Granularity& 0.1058& 0.0020& 0.1146& 0.0018\\
        4 Granularity& 0.1062& 0.0027& 0.1149& 0.0022\\
    \bottomrule
    \end{tabular}
\end{table}

\subsection{Multi-granularity Extension Study}
\label{sec:extension}
We further test whether using $3$ or $4$ granularity levels improves DIGCN. Each configuration is regenerated 5 times, and Table~\ref{tab:extension} reports the mean Recall@20 and standard deviation.

Additional granularities bring marginal gains but also larger standard deviations, likely because LLMs become less stable when asked to control multiple abstraction levels. We therefore keep the 2-level design as the default and view higher-granularity versions as optional choices for resource-rich settings.

\begin{figure}[!t]
    \centering
    \includegraphics[width=1\linewidth]{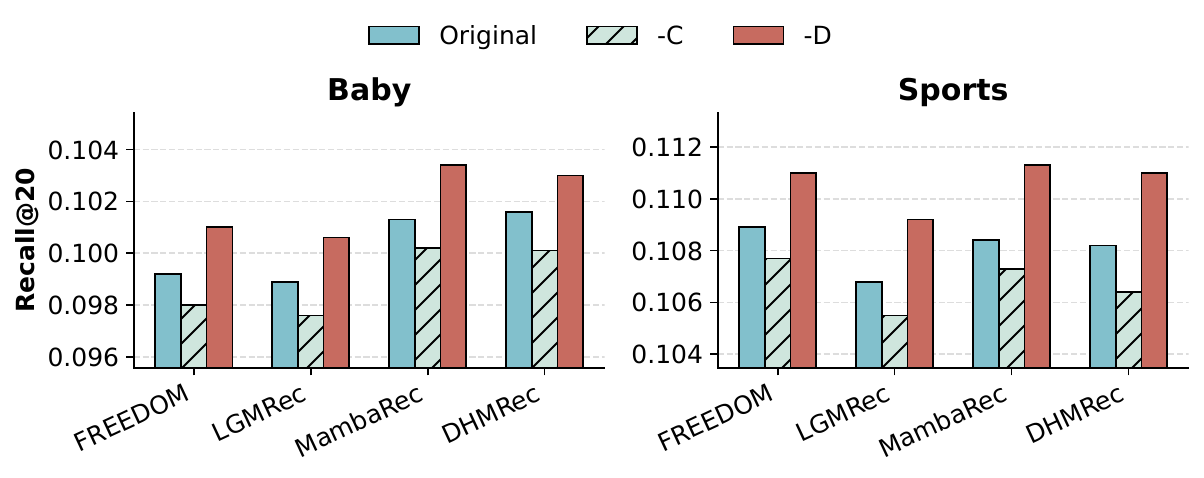}
    \caption{Recall@20 control for generated information. -C uses generated-only inputs, and -D uses dual inputs.}
    \label{fig:generation}
\end{figure}

\subsection{Analysis of Generated Information}
\label{sec:generation}
A potential concern is that DIGCN may benefit merely because LLM-generated content is better than the original features. We therefore compare representative strong baselines under three settings: Original, generated-only (-C), and dual-input (-D).

As shown in Figure~\ref{fig:generation}, -C generally degrades performance, indicating that generated coarse-grained information does not replace the original fine-grained features. The -D control further suggests that the useful signal comes from combining both granularities: coarse features capture general semantics, while original features retain detailed discriminative information.

\begin{figure}[!t]
    \centering
    \includegraphics[width=1\linewidth]{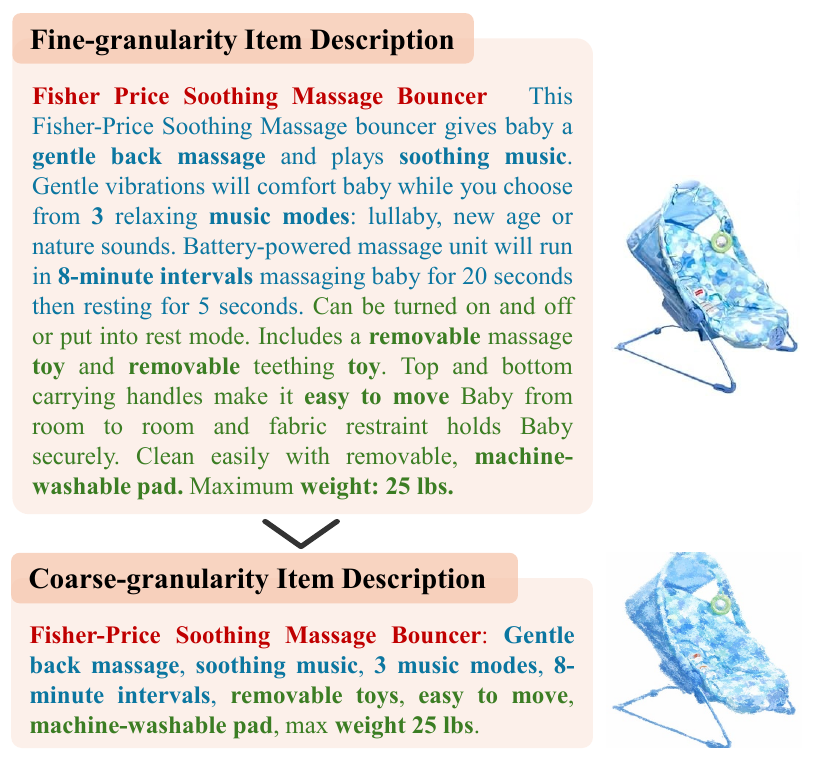}
    \caption{Case study on the Baby dataset.}
    \label{fig:case}
\end{figure}

\subsection{Case Study}
\label{sec:case}
To illustrate how dual-granularity raw features complement each other, Figure~\ref{fig:case} shows a Baby example. Same-color text denotes semantically equivalent content, bold words denote exact lexical matches, and underlined phrases denote concise synonym substitutions. The coarse-grained description is much shorter, preserves category-level cues such as "Gentle back massage" and "machine-washable pad", and removes overly technical details such as the precise massage interval. This increases semantic density without discarding the main item identity. The visual example follows the same pattern: the coarse-grained image weakens texture-level details while emphasizing color, shape, and structure. Thus, fine-grained features help distinguish similar items, whereas coarse-grained features provide broader semantic cues for category-level preference modeling.


\section{Conclusion}
\label{sec:conclusion}
This work addresses two key limitations in multimodal recommendation: low-quality representation learning and ineffective modality fusion. We propose DIGCN, a framework that (1) leverages LLMs to construct dual-granularity raw features before training, reducing the difficulty of multi-granularity learning from sparse data, and (2) introduces an intermediate fusion GCN with dynamic cross-modality weight correction for effective fusion. Extensive experiments show that DIGCN consistently outperforms advanced baselines and remains robust under data sparsity and cold-start scenarios. DIGCN is extensible and can benefit from additional alignment tasks.

\begin{acks}
This work was supported by the UGC General Research Fund no. 17209822 and the Innovation and Technology Commission Fund no. ITS/383/23FP from Hong Kong.
\end{acks}


\subsection*{Ethical Considerations}
This work uses publicly available recommendation datasets and does not involve collecting new user data or identifying individual users. The proposed dual-granularity raw feature construction relies on generative models to produce coarse-grained textual and visual references. Although these generated features are used only as auxiliary item-side information, they may inherit biases or inaccuracies from the underlying generators. We therefore keep the original item features as the fine-grained source, use generated information only as a complementary view, and empirically examine whether the improvement comes from dual-granularity modeling rather than replacing original information with generated content. In practical deployments, generated item features should be checked for factual consistency, harmful content, and domain-specific fairness concerns before being used in recommendation services.

\bibliographystyle{ACM-Reference-Format}
\bibliography{reference}

@article{zhou2023mmrecsm,
  author = {Zhou, Xin},
  title = {MMRec: Simplifying Multimodal Recommendation},
  year = {2023},
  journal={arXiv preprint arXiv:2302.03497},
}

@inproceedings{zhou2023tale,
  title={A tale of two graphs: Freezing and denoising graph structures for multimodal recommendation},
  author={Zhou, Xin and Shen, Zhiqi},
  booktitle={Proceedings of the 31st ACM International Conference on Multimedia},
  pages={935--943},
  year={2023}
}

@inproceedings{zhang2021mining,
  title={Mining latent structures for multimedia recommendation},
  author={Zhang, Jinghao and Zhu, Yanqiao and Liu, Qiang and Wu, Shu and Wang, Shuhui and Wang, Liang},
  booktitle={Proceedings of the 29th ACM International Conference on Multimedia},
  pages={3872--3880},
  year={2021}
}

@inproceedings{zhou2023bootstrap,
  title={Bootstrap latent representations for multi-modal recommendation},
  author={Zhou, Xin and Zhou, Hongyu and Liu, Yong and Zeng, Zhiwei and Miao, Chunyan and Wang, Pengwei and You, Yuan and Jiang, Feijun},
  booktitle={Proceedings of the ACM Web Conference 2023},
  pages={845--854},
  year={2023}
}

@inproceedings{he2020lightgcn,
  title={Lightgcn: Simplifying and powering graph convolution network for recommendation},
  author={He, Xiangnan and Deng, Kuan and Wang, Xiang and Li, Yan and Zhang, Yongdong and Wang, Meng},
  booktitle={Proceedings of the 43rd International ACM SIGIR conference on research and development in Information Retrieval},
  pages={639--648},
  year={2020}
}

@article{rendle2012bpr,
  title={BPR: Bayesian personalized ranking from implicit feedback},
  author={Rendle, Steffen and Freudenthaler, Christoph and Gantner, Zeno and Schmidt-Thieme, Lars},
  journal={arXiv preprint arXiv:1205.2618},
  year={2012}
}

@inproceedings{mcauley2015image,
  title={Image-based recommendations on styles and substitutes},
  author={McAuley, Julian and Targett, Christopher and Shi, Qinfeng and Van Den Hengel, Anton},
  booktitle={Proceedings of the 38th international ACM SIGIR conference on research and development in information retrieval},
  pages={43--52},
  year={2015}
}

@inproceedings{he2016vbpr,
  title={VBPR: visual bayesian personalized ranking from implicit feedback},
  author={He, Ruining and McAuley, Julian},
  booktitle={Proceedings of the AAAI conference on artificial intelligence},
  volume={30},
  number={1},
  year={2016}
}

@inproceedings{wei2019mmgcn,
  title={MMGCN: Multi-modal graph convolution network for personalized recommendation of micro-video},
  author={Wei, Yinwei and Wang, Xiang and Nie, Liqiang and He, Xiangnan and Hong, Richang and Chua, Tat-Seng},
  booktitle={Proceedings of the 27th ACM international conference on multimedia},
  pages={1437--1445},
  year={2019}
}

@article{tao2022self,
  title={Self-supervised learning for multimedia recommendation},
  author={Tao, Zhulin and Liu, Xiaohao and Xia, Yewei and Wang, Xiang and Yang, Lifang and Huang, Xianglin and Chua, Tat-Seng},
  journal={IEEE Transactions on Multimedia},
  year={2022},
  publisher={IEEE}
}

@article{wang2021dualgnn,
  title={Dualgnn: Dual graph neural network for multimedia recommendation},
  author={Wang, Qifan and Wei, Yinwei and Yin, Jianhua and Wu, Jianlong and Song, Xuemeng and Nie, Liqiang},
  journal={IEEE Transactions on Multimedia},
  year={2021},
  publisher={IEEE}
}

@inproceedings{glorot2010understanding,
  title={Understanding the difficulty of training deep feedforward neural networks},
  author={Glorot, Xavier and Bengio, Yoshua},
  booktitle={Proceedings of the thirteenth international conference on artificial intelligence and statistics},
  pages={249--256},
  year={2010},
  organization={JMLR Workshop and Conference Proceedings}
}

@article{kingma2014adam,
  title={Adam: A method for stochastic optimization},
  author={Kingma, Diederik P and Ba, Jimmy},
  journal={arXiv preprint arXiv:1412.6980},
  year={2014}
}

@inproceedings{wei2023multi,
  title={Multi-Modal Self-Supervised Learning for Recommendation},
  author={Wei, Wei and Huang, Chao and Xia, Lianghao and Zhang, Chuxu},
  booktitle={Proceedings of the ACM Web Conference 2023},
  pages={790--800},
  year={2023}
}

@inproceedings{xu2025mentor,
  title={Mentor: multi-level self-supervised learning for multimodal recommendation},
  author={Xu, Jinfeng and Chen, Zheyu and Yang, Shuo and Li, Jinze and Wang, Hewei and Ngai, Edith CH},
  booktitle={Proceedings of the AAAI Conference on Artificial Intelligence},
  volume={39},
  number={12},
  pages={12908--12917},
  year={2025}
}

@inproceedings{guo2024lgmrec,
  title={LGMRec: Local and Global Graph Learning for Multimodal Recommendation},
  author={Guo, Zhiqiang and Li, Jianjun and Li, Guohui and Wang, Chaoyang and Shi, Si and Ruan, Bin},
  booktitle={Proceedings of the AAAI Conference on Artificial Intelligence},
  volume={38},
  number={8},
  pages={8454--8462},
  year={2024}
}

@article{jiang2024diffmm,
  title={DiffMM: Multi-Modal Diffusion Model for Recommendation},
  author={Jiang, Yangqin and Xia, Lianghao and Wei, Wei and Luo, Da and Lin, Kangyi and Huang, Chao},
  journal={Proceedings of the 32nd ACM International Conference on Multimedia},
  year={2024}
}

@inproceedings{wei2024llmrec,
  title={Llmrec: Large language models with graph augmentation for recommendation},
  author={Wei, Wei and Ren, Xubin and Tang, Jiabin and Wang, Qinyong and Su, Lixin and Cheng, Suqi and Wang, Junfeng and Yin, Dawei and Huang, Chao},
  booktitle={Proceedings of the 17th ACM International Conference on Web Search and Data Mining},
  pages={806--815},
  year={2024}
}

@inproceedings{fan2025befa,
  title={BeFA: A General Behavior-driven Feature Adapter for Multimedia Recommendation},
  author={Fan, Qile and Yu, Penghang and Tan, Zhiyi and Bao, Bing-Kun and Lu, Guanming},
  booktitle={Proceedings of the AAAI Conference on Artificial Intelligence},
  volume={39},
  number={11},
  pages={11634--11644},
  year={2025}
}

@article{xu2024fourierkan,
  title={FourierKAN-GCF: Fourier Kolmogorov-Arnold Network--An Effective and Efficient Feature Transformation for Graph Collaborative Filtering},
  author={Xu, Jinfeng and Chen, Zheyu and Li, Jinze and Yang, Shuo and Wang, Wei and Hu, Xiping and Ngai, Edith C-H},
  journal={arXiv preprint arXiv:2406.01034},
  year={2024}
}

@inproceedings{zhan2026dhmrec,
  title={DHMRec: Collaboration-Guided Multimodal Disentanglement and Hierarchical Fusion for Recommendation},
  author={Zhan, Xiaohan and Shi, Yuliang and Wang, Jihu and Liu, Shijun and Kong, Fanyu and Chen, Zhiyong},
  booktitle={Proceedings of the AAAI Conference on Artificial Intelligence},
  volume={40},
  number={19},
  pages={16244--16252},
  year={2026}
}

@article{zhou2023comprehensive,
  title={A comprehensive survey on multimodal recommender systems: Taxonomy, evaluation, and future directions},
  author={Zhou, Hongyu and Zhou, Xin and Zeng, Zhiwei and Zhang, Lingzi and Shen, Zhiqi},
  journal={arXiv preprint arXiv:2302.04473},
  year={2023}
}

@inproceedings{xu2025nlgcl,
  title={NLGCL: Naturally Existing Neighbor Layers Graph Contrastive Learning for Recommendation},
  author={Xu, Jinfeng and Chen, Zheyu and Yang, Shuo and Li, Jinze and Wang, Hewei and Wang, Wei and Hu, Xiping and Ngai, Edith},
  booktitle={Proceedings of the Nineteenth ACM Conference on Recommender Systems},
  pages={319--329},
  year={2025}
}

@inproceedings{xu2025best,
  title={The Best is Yet to Come: Graph Convolution in the Testing Phase for Multimodal Recommendation},
  author={Xu, Jinfeng and Chen, Zheyu and Yang, Shuo and Li, Jinze and Ngai, Edith CH},
  booktitle={Proceedings of the 33rd ACM International Conference on Multimedia},
  pages={6325--6334},
  year={2025}
}

@inproceedings{chen2025don,
  title={Don’t Lose Yourself: Boosting Multimodal Recommendation via Reducing Node-neighbor Discrepancy in Graph Convolutional Network},
  author={Chen, Zheyu and Xu, Jinfeng and Hu, Haibo},
  booktitle={ICASSP 2025-2025 IEEE International Conference on Acoustics, Speech and Signal Processing (ICASSP)},
  pages={1--5},
  year={2025},
  organization={IEEE}
}

@article{xu2025vi,
  title={VI-MMRec: Similarity-Aware Training Cost-free Virtual User-Item Interactions for Multimodal Recommendation},
  author={Xu, Jinfeng and Chen, Zheyu and Yang, Shuo and Li, Jinze and Wan, Zitong and Wang, Hewei and Liu, Weijie and Li, Yijie and Ngai, Edith CH},
  journal={arXiv preprint arXiv:2512.08702},
  year={2025}
}

@inproceedings{xu2025cohesion,
  title={COHESION: Composite Graph Convolutional Network with Dual-Stage Fusion for Multimodal Recommendation},
  author={Xu, Jinfeng and Chen, Zheyu and Wang, Wei and Hu, Xiping and Kim, Sang-Wook and Ngai, Edith CH},
  booktitle={Proceedings of the 48th International ACM SIGIR Conference on Research and Development in Information Retrieval},
  pages={1830--1839},
  year={2025}
}

@inproceedings{xu2025mdvt,
  title={Mdvt: Enhancing multimodal recommendation with model-agnostic multimodal-driven virtual triplets},
  author={Xu, Jinfeng and Chen, Zheyu and Li, Jinze and Yang, Shuo and Wang, Hewei and Li, Yijie and Li, Mengran and Wu, Puzhen and Ngai, Edith CH},
  booktitle={Proceedings of the 31st ACM SIGKDD Conference on Knowledge Discovery and Data Mining V. 2},
  pages={3378--3389},
  year={2025}
}

@article{xu2025enhancing,
  title={Enhancing Robustness and Generalization Capability for Multimodal Recommender Systems via Sharpness-Aware Minimization},
  author={Xu, Jinfeng and Chen, Zheyu and Li, Jinze and Yang, Shuo and Wang, Wei and Hu, Xiping and Wong, Raymond Chi-Wing and Ngai, Edith CH},
  journal={IEEE Transactions on Knowledge and Data Engineering},
  year={2025},
  publisher={IEEE}
}

@article{xu2025survey,
  title={A Survey on Multimodal Recommender Systems: Recent Advances and Future Directions},
  author={Xu, Jinfeng and Chen, Zheyu and Yang, Shuo and Li, Jinze and Wang, Wei and Hu, Xiping and Hoi, Steven and Ngai, Edith},
  journal={arXiv preprint arXiv:2502.15711},
  year={2025}
}

@article{xu2025lobster,
  title={LOBSTER: Bilateral Global Semantic Enhancement for Multimedia Recommendation},
  author={Xu, Jinfeng and Chen, Zheyu and Wang, Wei and Hu, Xiping and Liu, Jiyi and Ngai, Edith CH},
  journal={Information Fusion},
  pages={103778},
  year={2025},
  publisher={Elsevier}
}

@inproceedings{chen2025hypercomplex,
  title={Hypercomplex Prompt-aware Multimodal Recommendation},
  author={Chen, Zheyu and Xu, Jinfeng and Wang, Hewei and Yang, Shuo and Wan, Zitong and Hu, Haibo},
  booktitle={Proceedings of the 34th ACM International Conference on Information and Knowledge Management},
  pages={403--414},
  year={2025}
}

@inproceedings{ren2025modality,
  title={Modality Alignment with Multi-scale Bilateral Attention for Multimodal Recommendation},
  author={Ren, Kelin and Ju, Chan-Yang and Lee, Dong-Ho},
  booktitle={Proceedings of the 34th ACM International Conference on Information and Knowledge Management},
  pages={2493--2502},
  year={2025}
}

@article{zhang2022latent,
  title={Latent structure mining with contrastive modality fusion for multimedia recommendation},
  author={Zhang, Jinghao and Zhu, Yanqiao and Liu, Qiang and Zhang, Mengqi and Wu, Shu and Wang, Liang},
  journal={IEEE Transactions on Knowledge and Data Engineering},
  volume={35},
  number={9},
  pages={9154--9167},
  year={2022},
  publisher={IEEE}
}

@inproceedings{xu2024aligngroup,
  title={AlignGroup: Learning and Aligning Group Consensus with Member Preferences for Group Recommendation},
  author={Xu, Jinfeng and Chen, Zheyu and Li, Jinze and Yang, Shuo and Wang, Hewei and Ngai, Edith CH},
  booktitle={Proceedings of the 33rd ACM International Conference on Information and Knowledge Management},
  pages={2682--2691},
  year={2024}
}

@inproceedings{chen2025squeeze,
  title={Squeeze and Excitation: A Weighted Graph Contrastive Learning for Collaborative Filtering},
  author={Chen, Zheyu and Xu, Jinfeng and Wei, Yutong and Peng, Ziyue},
  booktitle={Proceedings of the 48th International ACM SIGIR Conference on Research and Development in Information Retrieval},
  pages={2769--2773},
  year={2025}
}

@article{guo2025deepseek,
  title={Deepseek-r1 incentivizes reasoning in llms through reinforcement learning},
  author={Guo, Daya and Yang, Dejian and Zhang, Haowei and Song, Junxiao and Wang, Peiyi and Zhu, Qihao and Xu, Runxin and Zhang, Ruoyu and Ma, Shirong and Bi, Xiao and others},
  journal={Nature},
  volume={645},
  number={8081},
  pages={633--638},
  year={2025},
  publisher={Nature Publishing Group UK London}
}

@article{yang2025qwen3,
  title={Qwen3 technical report},
  author={Yang, An and Li, Anfeng and Yang, Baosong and Zhang, Beichen and Hui, Binyuan and Zheng, Bo and Yu, Bowen and Gao, Chang and Huang, Chengen and Lv, Chenxu and others},
  journal={arXiv preprint arXiv:2505.09388},
  year={2025}
}

@inproceedings{chen2019personalized,
  title={Personalized fashion recommendation with visual explanations based on multimodal attention network: Towards visually explainable recommendation},
  author={Chen, Xu and Chen, Hanxiong and Xu, Hongteng and Zhang, Yongfeng and Cao, Yixin and Qin, Zheng and Zha, Hongyuan},
  booktitle={Proceedings of the 42nd International ACM SIGIR Conference on Research and Development in Information Retrieval},
  pages={765--774},
  year={2019}
}

@inproceedings{yu2023multi,
  title={Multi-view graph convolutional network for multimedia recommendation},
  author={Yu, Penghang and Tan, Zhiyi and Lu, Guanming and Bao, Bing-Kun},
  booktitle={Proceedings of the 31st ACM International Conference on Multimedia},
  pages={6576--6585},
  year={2023}
}

\end{document}